\documentclass{svmult}
\usepackage{epsfig,graphicx} % figures
\usepackage[bottom]{footmisc}% places footnotes at page bottom
\usepackage{natbib,url}      %RR additions
\bibpunct{(}{)}{;}{a}{}{,}   %RR reference punctuation
\def\cite#1{\citealp{#1}}    %RR restore old astroncite \cite command
\def\authorindex#1{}  %RA to be redefined by editor at insertion into book

\begin{document}\newcount\preprintheader\preprintheader=1

%%\setcounter{page}{1}
%RA to insert and reset to actual page number for your Astro-PH upload

%RR file: rr-assp-defs.tex = extra ASSP definitions by Rob Rutten
%RR last: Feb 24 2009 
%RR note: %RR Rob-to-Rob    
%RR site: cp rr-assp-defs.tex ~/rr/tex/macros/.

\def\thisvolume{these proceedings}
%RR OOPS: no volume number and no page numbers, Springer muckup

%RR journal abbreviations
%%%%%%%%%%%%%%%%%%%%%%%%%
\def\aj{{AJ}}			
\def\araa{{ARA\&A}}		
\def\apj{{ApJ}}			
\def\apjl{{ApJ}}		
\def\apjs{{ApJS}}		
\def\ao{{Appl.\ Optics}} 
\def\apss{{Ap\&SS}}		
\def\aap{{A\&A}}		
\def\aapr{{A\&A~Rev.}}		
\def\aaps{{A\&AS}}		
\def\an{{Astron.\ Nachrichten}}
\def\aspcs{{ASP Conf.\ Ser.}}
\def\assp{{Astrophys.\ \& Space Sci.\ Procs., Springer, Heidelberg}}
\def\azh{{AZh}}			
\def\baas{{BAAS}}		
\def\jrasc{{JRASC}}	
\def\memras{{MmRAS}}		
\def\mnras{{MNRAS}}
\def\nat{{Nat}}		
\def\pra{{Phys.\ Rev.\ A}} 
\def\prb{{Phys.\ Rev.\ B}}		
\def\prc{{Phys.\ Rev.\ C}}		
\def\prd{{Phys.\ Rev.\ D}}		
\def\prl{{Phys.\ Rev.\ Lett.}} %RR	
\def\pasp{{PASP}}
\def\pasj{{PASJ}}		
\def\qjras{{QJRAS}}
\def\science{{Sci}}		
\def\skytel{{S\&T}}		
\def\solphys{{Solar\ Phys.}} 
\def\sovast{{Soviet\ Ast.}}  
\def\ssr{{Space\ Sci.\ Rev.}}
\def\svassp{{Astrophys.\ Space Sci.\ Procs., Springer, Heidelberg}}
\def\zap{{ZAp}}			
\let\astap=\aap
\let\apjlett=\apjl
\let\apjsupp=\apjs
\def\grl{{Geophys.\ Res.\ Lett.}}  %RR Weiss
\def\jgr{{J. Geophys.\ Res.}} %RR Manoharan

%RR astronomy and math commands copied from ASP
%%%%%%%%%%%%%%%%%%%%%%%%%%%%%%%%%%%%%%%%%%%%%%%
\def\ion#1#2{{\rm #1}\,{\uppercase{#2}}}  %RR ~>\, \sc > uc 
\def\deg{\hbox{$^\circ$}}
\def\sun{\hbox{$\odot$}}
\def\earth{\hbox{$\oplus$}}
\def\la{\mathrel{\hbox{\rlap{\hbox{\lower4pt\hbox{$\sim$}}}\hbox{$<$}}}}
\def\ga{\mathrel{\hbox{\rlap{\hbox{\lower4pt\hbox{$\sim$}}}\hbox{$>$}}}}
\def\sq{\hbox{\rlap{$\sqcap$}$\sqcup$}}
\def\arcmin{\hbox{$^\prime$}}
\def\arcsec{\hbox{$^{\prime\prime}$}}
\def\fd{\hbox{$.\!\!^{\rm d}$}}
\def\fh{\hbox{$.\!\!^{\rm h}$}}
\def\fm{\hbox{$.\!\!^{\rm m}$}}
\def\fs{\hbox{$.\!\!^{\rm s}$}}
\def\fdg{\hbox{$.\!\!^\circ$}}
\def\farcm{\hbox{$.\mkern-4mu^\prime$}}
\def\farcs{\hbox{$.\!\!^{\prime\prime}$}}
\def\fp{\hbox{$.\!\!^{\scriptscriptstyle\rm p}$}}
\def\micron{\hbox{$\mu$m}}
\def\onehalf{\hbox{$\,^1\!/_2$}}	
\def\onethird{\hbox{$\,^1\!/_3$}}
\def\twothirds{\hbox{$\,^2\!/_3$}}
\def\onequarter{\hbox{$\,^1\!/_4$}}
\def\threequarters{\hbox{$\,^3\!/_4$}}
\def\ubv{\hbox{$U\!BV$}}		
\def\ubvr{\hbox{$U\!BV\!R$}}		
\def\ubvri{\hbox{$U\!BV\!RI$}}		
\def\ubvrij{\hbox{$U\!BV\!RI\!J$}}		
\def\ubvrijh{\hbox{$U\!BV\!RI\!J\!H$}}		
\def\ubvrijhk{\hbox{$U\!BV\!RI\!J\!H\!K$}}		
\def\ub{\hbox{$U\!-\!B$}}		
\def\bv{\hbox{$B\!-\!V$}}		
\def\vr{\hbox{$V\!-\!R$}}		
\def\ur{\hbox{$U\!-\!R$}}

%%%%%%%%%%%%%%%%%%%%%%%%%%%%%%%%%%%%%%%%%%%%%%%%%%%%%%%%%%%%%%%%%%%%%%%%%%%%
%RR RJR additional commands
%%%%%%%%%%%%%%%%%%%%%%%%%%%%%%%%%%%%%%%%%%%%%%%%%%%%%%%%%%%%%%%%%%%%%%%%%%%%

%RR -- non-bullet item marker in itemize list 
\def\labelitemi{{\bf --}}  

%RR -- latin abbreviations
\def\rmit#1{{\it #1}}              %% italics (RR style, Kluwer)
\def\rmit#1{{\rm #1}}              %% redefine for ASP, A&A, ApJ, Springer
\def\etal{\rmit{et al.}}           %% use \etal\ for space behind it        
\def\etc{\rmit{etc.}}           
\def\ie{\rmit{i.e.,}}              %% , required (Webster 1681)
\def\eg{\rmit{e.g.,}}              %% , required (Webster 1681)
\def\cf{cf.}                       %% no Latin, always Roman (Webster 1686)
\def\viz{\rmit{viz.}}
\def\vs{\rmit{vs.}}

%RR -- mathematical
\def\rot{\hbox{\rm rot}}
\def\div{\hbox{\rm div}}
\def\lesssim{\mathrel{\hbox{\rlap{\hbox{\lower4pt\hbox{$\sim$}}}\hbox{$<$}}}}
\def\gtrsim{\mathrel{\hbox{\rlap{\hbox{\lower4pt\hbox{$\sim$}}}\hbox{$>$}}}}
\def\dif{\: {\rm d}}                       %% differential d with space
\def\ep{\:{\rm e}^}                        %% e^ with space and roman e
\def\dash{\hbox{$\,-\,$}}                  %% math-like hyphen
\def\is{\!=\!}                             %% = in text for tighter spacing

%RR --stellar stuff
\def\starname#1#2{${#1}$\,{\rm {#2}}}  %% \starname{\alpha}{Cen~A} 
\def\Teff{\hbox{$T_{\rm eff}$}}   

%RR -- units (in addition to the ASP ones above)
\def\kms{\hbox{km$\;$s$^{-1}$}}
\def\ms{\hbox{m$\;$s$^{-1}$}}
\def\Mxcm{\hbox{Mx\,cm$^{-2}$}}    %% no 2, damn tex

%RR -- magnetic field 
\def\Bapp{\hbox{$B_{\rm app}$}}    %% apparent flux density, Lites convention

%RR -- oscillations
\def\komega{($k, \omega$)}                 %% k - omega 
\def\kf{($k_h,f$)}                         %% f - k_h
\def\VminI{\hbox{$V\!\!-\!\!I$}}           %% V-I
\def\IminI{\hbox{$I\!\!-\!\!I$}}           %% I-I
\def\VminV{\hbox{$V\!\!-\!\!V$}}           %% V-V
\def\Xt{\hbox{$X\!\!-\!t$}}                %% X-t

%RR -- atomic levels
%%      use:    \level 3s3p 3Pe
%%              \level 3s$^2$ {1,3}P{e,o}
%%              \level {} 3Ge
\def\level #1 #2#3#4{$#1 \: ^{#2} \mbox{#3} ^{#4}$}   

%RR -- some spectral species
\def\specchar#1{\uppercase{#1}}    %% to be redefined for A&A = \sc
\def\AlI{\mbox{Al\,\specchar{i}}}  %% use \AlI\ for space behind it
\def\BI{\mbox{B\,\specchar{i}}} 
\def\BII{\mbox{B\,\specchar{ii}}}  
\def\BaI{\mbox{Ba\,\specchar{i}}}  
\def\BaII{\mbox{Ba\,\specchar{ii}}} 
\def\CI{\mbox{C\,\specchar{i}}} 
\def\CII{\mbox{C\,\specchar{ii}}} 
\def\CIII{\mbox{C\,\specchar{iii}}} 
\def\CIV{\mbox{C\,\specchar{iv}}} 
\def\CaI{\mbox{Ca\,\specchar{i}}} 
\def\CaII{\mbox{Ca\,\specchar{ii}}} 
\def\CaIII{\mbox{Ca\,\specchar{iii}}} 
\def\CoI{\mbox{Co\,\specchar{i}}} 
\def\CrI{\mbox{Cr\,\specchar{i}}} 
\def\CriI{\mbox{Cr\,\specchar{ii}}} 
\def\CsI{\mbox{Cs\,\specchar{i}}} 
\def\CsII{\mbox{Cs\,\specchar{ii}}} 
\def\CuI{\mbox{Cu\,\specchar{i}}} 
\def\FeI{\mbox{Fe\,\specchar{i}}} 
\def\FeII{\mbox{Fe\,\specchar{ii}}} 
\def\FeIX{\mbox{Fe\,\specchar{ix}}}
\def\FeX{\mbox{Fe\,\specchar{x}}}
\def\FeXVI{\mbox{Fe\,\specchar{xvi}}}
\def\FrI{\mbox{Fr\,\specchar{i}}}
\def\HI{\mbox{H\,\specchar{i}}} 
\def\HII{\mbox{H\,\specchar{ii}}} 
\def\Hmin{\hbox{\rmH$^{^{_{\scriptstyle -}}}$}}      %% H^min, elegant
\def\Hemin{\hbox{{\rm He}$^{^{_{\scriptstyle -}}}$}} %% He^min, idem
\def\HeI{\mbox{He\,\specchar{i}}} 
\def\HeII{\mbox{He\,\specchar{ii}}} 
\def\HeIII{\mbox{He\,\specchar{iii}}} 
\def\KI{\mbox{K\,\specchar{i}}} 
\def\KII{\mbox{K\,\specchar{ii}}} 
\def\KIII{\mbox{K\,\specchar{iii}}} 
\def\LiI{\mbox{Li\,\specchar{i}}} 
\def\LiII{\mbox{Li\,\specchar{ii}}} 
\def\LiIII{\mbox{Li\,\specchar{iii}}} 
\def\MgI{\mbox{Mg\,\specchar{i}}} 
\def\MgII{\mbox{Mg\,\specchar{ii}}} 
\def\MgIII{\mbox{Mg\,\specchar{iii}}} 
\def\MnI{\mbox{Mn\,\specchar{i}}} 
\def\NI{\mbox{N\,\specchar{i}}}
\def\NIV{\mbox{N\,\specchar{iv}}}
\def\NaI{\mbox{Na\,\specchar{i}}}
\def\NaII{\mbox{Na\,\specchar{ii}}}
\def\NaIII{\mbox{Na\,\specchar{iii}}}
\def\NeVIII{\mbox{Ne\,\specchar{viii}}} 
\def\NiI{\mbox{Ni\,\specchar{i}}} 
\def\NiII{\mbox{Ni\,\specchar{ii}}}
\def\NiIII{\mbox{Ni\,\specchar{iii}}} 
\def\OI{\mbox{O\,\specchar{i}}} 
\def\OVI{\mbox{O\,\specchar{vi}}}
\def\RbI{\mbox{Rb\,\specchar{i}}} 
\def\SII{\mbox{S\,\specchar{ii}}} 
\def\SiI{\mbox{Si\,\specchar{i}}} 
\def\SiII{\mbox{Si\,\specchar{ii}}} 
\def\SrI{\mbox{Sr\,\specchar{i}}}
\def\SrII{\mbox{Sr\,\specchar{ii}}}
\def\TiI{\mbox{Ti\,\specchar{i}}} 
\def\TiII{\mbox{Ti\,\specchar{ii}}} 
\def\TiIII{\mbox{Ti\,\specchar{iii}}} 
\def\TiIV{\mbox{Ti\,\specchar{iv}}} 
\def\VI{\mbox{V\,\specchar{i}}} 
\def\HtwoO{\mbox{H$_2$O}}        %% H2O %RR TeX doesn't accept numbers alas
\def\Otwo{\mbox{O$_2$}}          %% O2

%RR -- hydrogen spectrum features
\def\Halpha{\mbox{H\hspace{0.1ex}$\alpha$}} %% \Halpha\ for space behind it
\def\Ha{\mbox{H\hspace{0.2ex}$\alpha$}}
\def\Hbeta{\mbox{H\hspace{0.2ex}$\beta$}}
\def\Hgamma{\mbox{H\hspace{0.2ex}$\gamma$}}
\def\Hdelta{\mbox{H\hspace{0.2ex}$\delta$}}
\def\Hepsilon{\mbox{H\hspace{0.2ex}$\epsilon$}}
\def\Hzeta{\mbox{H\hspace{0.2ex}$\zeta$}}
\def\Lyalpha{\mbox{Ly$\hspace{0.2ex}\alpha$}}
\def\Lybeta{\mbox{Ly$\hspace{0.2ex}\beta$}}
\def\Lygamma{\mbox{Ly$\hspace{0.2ex}\gamma$}}
\def\Lycont{\mbox{Ly\hspace{0.2ex}{\small cont}}}
\def\Baalpha{\mbox{Ba$\hspace{0.2ex}\alpha$}}
\def\Babeta{\mbox{Ba$\hspace{0.2ex}\beta$}}
\def\Bacont{\mbox{Ba\hspace{0.2ex}{\small cont}}}
\def\Paalpha{\mbox{Pa$\hspace{0.2ex}\alpha$}}
\def\Bralpha{\mbox{Br$\hspace{0.2ex}\alpha$}}

%RR -- Na D
\def\NaD{\mbox{Na\,\specchar{i}\,D}}    %% use \NaD\ for space behind it
\def\NaDone{\mbox{Na\,\specchar{i}\,\,D$_1$}}
\def\NaDtwo{\mbox{Na\,\specchar{i}\,\,D$_2$}}
\def\NaID{\mbox{Na\,\specchar{i}\,\,D}}
\def\NaIDone{\mbox{Na\,\specchar{i}\,\,D$_1$}}
\def\NaIDtwo{\mbox{Na\,\specchar{i}\,\,D$_2$}}
\def\Done{\mbox{D$_1$}}
\def\Dtwo{\mbox{D$_2$}}

%RR -- Mg b 
\def\Mgbone{\mbox{Mg\,\specchar{i}\,b$_1$}}
\def\Mgbtwo{\mbox{Mg\,\specchar{i}\,b$_2$}}
\def\Mgbthree{\mbox{Mg\,\specchar{i}\,b$_3$}}
\def\MgIb{\mbox{Mg\,\specchar{i}\,b}}
\def\MgIbone{\mbox{Mg\,\specchar{i}\,b$_1$}}
\def\MgIbtwo{\mbox{Mg\,\specchar{i}\,b$_2$}}
\def\MgIbthree{\mbox{Mg\,\specchar{i}\,b$_3$}}

%RR -- Ca II H & K 
\def\CaIIK{\mbox{Ca\,\specchar{ii}\,K}}       %% use \CaIIK\ for space
\def\CaIIH{\mbox{Ca\,\specchar{ii}\,H}}
\def\CaIIHK{\mbox{Ca\,\specchar{ii}\,H\,\&\,K}}
\def\HK{\mbox{H\,\&\,K}}
\def\Kthree{\mbox{K$_3$}}      %% numbers not permitted, alas
\def\Hthree{\mbox{H$_3$}}
\def\Ktwo{\mbox{K$_2$}}
\def\Htwo{\mbox{H$_2$}}
\def\Kone{\mbox{K$_1$}}     
\def\Hone{\mbox{H$_1$}}     
\def\KtwoV{\mbox{K$_{2V}$}}
\def\KtwoR{\mbox{K$_{2R}$}}
\def\KoneV{\mbox{K$_{1V}$}}
\def\KoneR{\mbox{K$_{1R}$}}
\def\HtwoV{\mbox{H$_{2V}$}}
\def\HtwoR{\mbox{H$_{2R}$}}
\def\HoneV{\mbox{H$_{1V}$}}
\def\HoneR{\mbox{H$_{1R}$}}

%RR -- Mg II h & k 
\def\hk{\mbox{h\,\&\,k}}
\def\kthree{\mbox{k$_3$}}    
\def\hthree{\mbox{h$_3$}}
\def\ktwo{\mbox{k$_2$}}
\def\htwo{\mbox{h$_2$}}
\def\kone{\mbox{k$_1$}}     
\def\hone{\mbox{h$_1$}}     
\def\ktwoV{\mbox{k$_{2V}$}}
\def\ktwoR{\mbox{k$_{2R}$}}
\def\koneV{\mbox{k$_{1V}$}}
\def\koneR{\mbox{k$_{1R}$}}
\def\htwoV{\mbox{h$_{2V}$}}
\def\htwoR{\mbox{h$_{2R}$}}
\def\honeV{\mbox{h$_{1V}$}}
\def\honeR{\mbox{h$_{1R}$}}

%%%%%%%%%%%%%%%%%%%%%%%%%%%%%%%%%%%%%%%%%%%%%%%%%%%%%%%%%% preprint header
%RR redefine @maketitle in svmult.cls to add slug on top
\ifnum\preprintheader=1     %RR ADAPT: 0 or 1 = preprintheader 
\makeatletter  %RR redefine symbol @ (trick from Pit Suetterlin)
\def\@maketitle{\newpage
\markboth{}{}%
%RR================================= Feb 27 2009 
  {\mbox{} \vspace*{-8ex} \par 
   \em \footnotesize To appear in ``Magnetic Coupling between the Interior 
       and the Atmosphere of the Sun'', eds. S.~S.~Hasan and R.~J.~Rutten, 
       Astrophysics and Space Science Proceedings, Springer-Verlag, 
       Heidelberg, Berlin, 2009.} \vspace*{-5ex} \par
%RR=================================
 \def\lastand{\ifnum\value{@inst}=2\relax
                 \unskip{} \andname\
              \else
                 \unskip \lastandname\
              \fi}%
 \def\and{\stepcounter{@auth}\relax
          \ifnum\value{@auth}=\value{@inst}%
             \lastand
          \else
             \unskip,
          \fi}%
  \raggedright
 {\Large \bfseries\boldmath
  \pretolerance=10000
  \let\\=\newline
% \@hangfrom{\@svsec}%
%%%  \@svsec
  \raggedright
  \hyphenpenalty \@M
  \interlinepenalty \@M
  \if@numart
     \chap@hangfrom{}%!!!
  \else
     \chap@hangfrom{\thechapter\thechapterend\hskip\betweenumberspace}%!!!
  \fi
  \ignorespaces
  \@title \par}\vskip .8cm
\if!\@subtitle!\else {\large \bfseries\boldmath
  \vskip -.65cm
  \pretolerance=10000
  \@subtitle \par}\vskip .8cm\fi
 \setbox0=\vbox{\setcounter{@auth}{1}\def\and{\stepcounter{@auth}}%
 \def\thanks##1{}\@author}%
 \global\value{@inst}=\value{@auth}%
 \global\value{auco}=\value{@auth}%
 \setcounter{@auth}{1}%
{\lineskip .5em
\noindent\ignorespaces
\@author\vskip.35cm}
 {\small\institutename\par}
 \ifdim\pagetotal>157\p@
     \vskip 11\p@
 \else
     \@tempdima=168\p@\advance\@tempdima by-\pagetotal
     \vskip\@tempdima
 \fi
}
\makeatother     %RR define @ back
\fi

%RA use these as needed but don't change nor add!
%RA   if you really need private /def's then define them here
%RA   so that we can inspect them and put into rr-assp-defs
%RA   if really useful - and not upsetting any other paper

\title*{Signatures of Coronal Heating Mechanisms}
%RA capitalize the nouns, try to fit on one line

\titlerunning{Coronal Heating Mechanisms}
%RA capitalize the nouns
%RA only if needed for a too long title, not in this case

\author{P. Antolin\inst{1,2} \and 
		K. Shibata\inst{1} \and 
		T. Kudoh\inst{3} \and 
		D. Shiota\inst{4} \and 
		D. Brooks\inst{5,6}}
%RA full initials but no first names; spaces between initials
%RA use \inst and \and, see rr-assp-example.tex 

\authorindex{Antolin, P.} 
\authorindex{Shibata, K.}
\authorindex{Kudoh, T.}
\authorindex{Shiota, D.}
\authorindex{Brooks, D.}
%RA please add, 1 line/author: {Name, A. B.} with spaces between initials

\authorrunning{Antolin et al.}
%RA insert this for three or more authors

\institute{Kwasan Observatory, Kyoto University, Japan 
		\and
		The Institute of Theoretical Astrophysics, University of Oslo, Norway
		\and
		National Astronomical Observatory of Japan, Japan
		\and The Earth Simulator Center, Japan Agency for Marine-Earth Science and Technology (JAMSTEC), Japan
		\and
		Space Science Division, Naval Research Laboratory, USA
		\and
		George Mason University, USA}
%RA no full postal addresses, just brief affiliation; no email addresses

\maketitle

\setcounter{footnote}{0}  %RR Springer forgot this one (and much more)

%%%%%%%%%%%%%%%%%%%%%%%%%%%%%%%%%%%%%%%%%%%%%%%%%%%%%%%%%%%%%%%%%%%%%%%%%%%%
\begin{abstract} 

Alfv\'en waves created by sub-photospheric motions or by magnetic
reconnection in the low solar atmosphere seem good candidates for
coronal heating. However, the corona is also likely to be heated more
directly by magnetic reconnection, with dissipation taking place in
current sheets. Distinguishing observationally between these two
heating mechanisms is an extremely difficult task.  We perform
1.5-dimensional MHD simulations of a coronal loop subject to each type
of heating and derive observational quantities that may allow these to
be differentiated. 
%RP twice observations 
This work is presented in more detail in
\citet{2008ApJ...688..669A}.

\end{abstract}

\section{Introduction}\label{antolin-sec:introduction}

The ``coronal heating problem'', \ie\ the heating of the solar corona
up to a few hundred times the average temperature of the underlying
photosphere, is one of the most perplexing and unresolved
problems in astrophysics to date.  Alfv\'en waves produced by the constant
turbulent convective motions or by magnetic reconnection in the lower
and upper solar atmosphere may transport enough energy to heat and
maintain a corona \citep{1974SoPh...35..451U}. A possible dissipation
mechanism for Alfv\'en waves is mode conversion. This is known as the
Alfv\'en wave heating model \citep{1982SoPh...75...35H,
1999ApJ...514..493K}. 

Another promising coronal heating mechanism is the nanoflare
reconnection heating model, first suggested by
\citet{1988ApJ...330..474P}, who considered coronal loops being
subject to many magnetic reconnection events, releasing energy
impulsively and sporadically in small quantities of the order of
$10^{24}$~erg or less (``nanoflares''), uniformly along loops. It has been
shown that both these candidate mechanisms can
account for the observed impulsive and ubiquitous character of
the heating events in the corona \citep{2001ApJ...557..343K, 2004ApJ...601L.107M}. 
How then can we distinguish observationally between both heating
mechanisms when these operate in the corona?

We propose 
%RP In this work - of course, take out
a way to discern observationally between Alfv\'en wave
heating and nanoflare reconnection heating. The idea relies on the
fact that the distribution of the shocks in loops differs
substantially between the two models, due to the different
characteristics of the wave modes they produce. As a consequence,
X-ray intensity profiles differ substantially between an Alfv\'en-wave
heated corona and a nanoflare-heated corona. The heating events
obtained follow a power-law distribution in frequency, with indices
which differ significantly from one heating model to the other. We
thus analyze the link between the power-law index of the frequency
distribution and the operating heating mechanism in the loop. We also
predict different flow structures and different average plasma
velocities along the loop, depending on the heating mechanism and its
spatial distribution.

\begin{figure}  
%RA no location specifier [t] or such please
  \centering
  \includegraphics[width=0.6\textwidth]{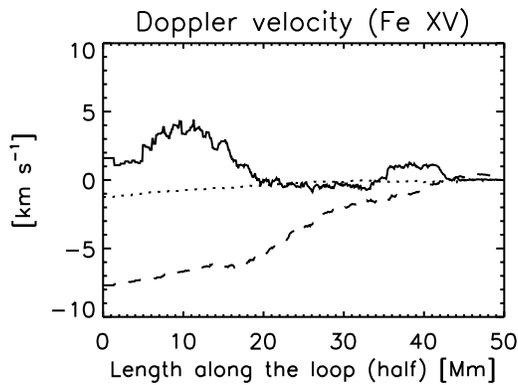}
%RA or smaller by declaring cm; the text width = 11.8cm 
  \caption[]{\label{antolin-fig:1}
%RA don't forget the [] 
%
  Doppler velocities from synthesis of Fe\,XV 284.16~\AA\
  emission with respect to distance along the first half of the
  loop. The loop is assumed to be at disk center.  The CHIANTI atomic
  database was used for calculation of the line profiles
  \citep{1997A&AS..125..149D, 2006ApJS..162..261L}. The solid, dashed
  and dotted curves correspond respectively to Alfv\'en wave heating,
%RP not lines, lines are straight
  footpoint nanoflare heating, and uniform nanoflare heating. Positive
  velocities correspond to redshifts (downflows).
}\end{figure}

\begin{table}
\centering
\caption[]{\label{antolin-table}
  Observational signatures for coronal heating mechanisms. We consider
  Alfv\'en wave heating, nanoflare-reconnection heating with heating
  events concentrated at the footpoints (``footpoint nanoflare'') or
  uniformly distributed in the corona (``uniform nanoflare''). Second
  column: mean and maximum flow velocities. Third column: Doppler
  velocities derived from synthesized Fe\,XV 284.16~\AA\ emission
  line. The loops is assumed to be observed from above (disk
%RP disk
  center). Fourth column: shape of the intensity flux time
  series. Fifth column: mean over the power law indexes obtained from
  intensity histograms for many positions along the loop from the
  footpoint to the apex.}
\begin{tabular*}{\textwidth}{@{\extracolsep{\fill}}cccccc}
\hline
Heating & Mean \& maximum & Doppler velocities & Intensity flux & Mean power \\
model & velocities [km\,s$^{-1}$] & (Fe\,XV) [km\,s$^{-1}$] & pattern & law index \\
\hline
Alfv\'en & $\langle v\rangle\sim 50$ & red shifts & bursty & $\langle\delta\rangle <$ -2 \\
wave & $v_{\rm max}> 200$ & $\sim$3 & everywhere & \\
\hline
Footpoint & $\langle v\rangle\sim$ 15 & blue shifts & bursty close & $-1.5 > \langle\delta\rangle> -2$ \\
nanoflare & $v_{\rm max}> 200$ & $\sim 7$ & to footpoints & decreases \\
\hline
Uniform & $\langle v\rangle\sim$ 5 & blue shifts & flat & $\langle\delta\rangle\sim -1$ \\
nanoflare & $v_{\rm max}<$ 40 & $\sim$ 1 & everywhere & decreases \\
\hline
\end{tabular*}
\end{table}
%RP  redid this according to Spinger authinst.pdf
%RP don't put $ before number

\section{Signatures for Alfv\'en wave heating}\label{antolin-sec:alfven}

Alfv\'en waves generated at the photosphere, due to nonlinear effects,
convert into longitudinal modes during propagation, with the major
conversion happening in the chromosphere. An important fraction of the
Alfv\'enic energy is also converted into slow and fast modes in the
corona, where the plasma $\beta$ parameter can get close to unity
sporadically and spontaneously. The resulting longitudinal modes
produce strong shocks which heat the plasma uniformly.  The result is
a uniform loop satisfying the RTV scaling law
\citep{1978ApJ...220..643R, 2004ApJ...601L.107M}, which is however
very dynamic (Table~\ref{antolin-table}). Synthetic Fe\,XV
emission lines show a predominance of red shifts (downflows) close to
the footpoints (Fig.~\ref{antolin-fig:1}). Synthetic XRT intensity
profiles show spiky patterns throughout the corona. Corresponding
intensity histograms show a distribution of heating events which stays
roughly constant along the corona, and which can be approximated by a
power law with index steeper than $-2$, an indication that most of the
heating comes from small dissipative events
\citep{1991SoPh..133..357H}.

\section{Signatures for nanoflare reconnection heating}\label{antolin-sec:nanoflare}

The nanoflare reconnection model consists of artificial injections of
energy, simulating nanoflares which can be distributed towards the
footpoints of the loop (``footpoint nanoflare heating'') or uniformly
along the corona (``uniform nanoflare heating''). Strong slow shocks
created by such heating events are only obtained in the first case,
close to the footpoints. Fast dissipation of the slow shocks by
thermal conduction leads to only weak shocks near the apex of the
loop. We only obtain weak shocks in the case of uniform nanoflare
heating. For both cases, the mean flow speeds are considerably smaller
than in the Alfv\'en wave heating model
(Table~\ref{antolin-table}). 
%RP don't use \cf everywhere
Synthetic Fe\,XV emission lines show,
mainly for footpoint heating, a predominance of blueshifts (upflows)
close to the footpoints (cf.\ Fig.~\ref{antolin-fig:1}) which may
match observations of active regions \citep{2008ApJ...678L..67H}. In
this case, spiky patterns result in XRT intensity profiles close to
the footpoints, and a flattening of the profile at the apex. The
corresponding power-law index of the heating events distribution
decreases closer to the apex. In the case of uniform nanoflare
heating, the intensity profiles are not spiky but rather uniform in
time due to the stronger dissipation by thermal conduction. It results
in a very shallow average power-law index for the distribution of
heating events.

\section{Conclusion}\label{antolin-sec:conclusion}

The observable differences between the two coronal heating mechanisms
are summarized in Table~\ref{antolin-table}. The power-law index of
the heating distribution is found to be sensible to the location of
the heating along coronal loops and to the heating mechanism
itself. Different flow patterns are also obtained.  Downflows of hot plasma
are present in the Alfv\'en-wave heating model, whereas hot upflows
are obtained for nanoflare reconnection heating. Footpoint nanoflare
heating seems to match the observations in active region
loops better.  Are Alfv\'en wave heating or uniform nanoflare heating 
more adequate for quiet-Sun loops?

%%%%%%%%%%%%%%%%%%%%%%%%%%%%%%%%%%%%%%%%%%%%%%%%%%%%%%%%%%%%%%%%%%%%%%%%%%%%
\begin{acknowledgement}
  We thank the conference organisers for a very good meeting and the
  editors for excellent instructions. %RA -:)
\end{acknowledgement}

%%%%%%%%%%%%%%%%%%%%%%%%%%%%%%%%%%%%%%%%%%%%%%%%%%%%%%%%%%%%%%%%%%%%%%%%%%%%
%% References
%%%%%%%%%%%%%%%%%%%%%%%%%%%%%%%%%%%%%%%%%%%%%%%%%%%%%%%%%%%%%%%%%%%%%%%%%%%%
\begin{small}

%RA Use commands as the following two to generate the bibliography
%RA automatically with BibTeX as file xxx.bbl.  

%%\bibliographystyle{rr-assp}       %RR hacked from aa.bst
%%\bibliography{/tmp/adsfiles.bib,bangalore.bib}  

\end{small}

\end{document}